\documentclass[aps,prd,twocolumn,superscriptaddress,nofootinbib]{revtex4-1}

\usepackage{graphicx,color}
\usepackage{latexsym,amsmath,amssymb,graphicx,booktabs}
\usepackage{hyperref}

\definecolor{MyBlue}{rgb}{0.15,0.15,0.70}
\hypersetup{
colorlinks=true,
citecolor=MyBlue,
linkcolor=MyBlue,
urlcolor=MyBlue
}

\newcommand{\dgw}{d_L^{\,\rm gw}}
\newcommand{\dem}{d_L^{\,\rm em}}

\newcommand{\nn}{\nonumber}

\renewcommand\({\left(}
\renewcommand\){\right)}
\renewcommand\[{\left[}
\renewcommand\]{\right]}

\def\lsim{\raise 0.4ex\hbox{$<$}\kern -0.8em\lower 0.62
ex\hbox{$\sim$}}

\def\gsim{\raise 0.4ex\hbox{$>$}\kern -0.7em\lower 0.62
ex\hbox{$\sim$}}

\def\lbar{{\hbox{$\lambda$}\kern -0.7em\raise 0.6ex
\hbox{$-$}}}

\newcommand\eq[1]{eq.~(\ref{#1})}

\newcommand\Eq[1]{Equation~(\ref{#1})}

\newcommand\eqst[2]{eqs.~(\ref{#1})--(\ref{#2})}

\newcommand\p{\partial}

\newcommand\ee{\end{equation}}
\newcommand\be{\begin{equation}}
\def\bea{\begin{array}}
\def\eea{\end{array}}\def\ea{\end{array}}
\newcommand\ees{\end{eqnarray}}
\newcommand\bees{\begin{eqnarray}}
\def\nn{\nonumber}

\def\dslash{\hspace{-1mm}\not{\hbox{\kern-2pt $\partial$}}}
\def\Dslash{\not{\hbox{\kern-2pt $D$}}}
\def\pslash{\not{\hbox{\kern-2.1pt $p$}}}
\def\kslash{\not{\hbox{\kern-2.3pt $k$}}}
\def\qslash{\not{\hbox{\kern-2.3pt $q$}}}

\def\p1{{\bf p}_1}
\def\p2{{\bf p}_2}
\def\k1{{\bf k}_1}
\def\k2{{\bf k}_2}

\newcommand{\dddM}{\kern 0.2em \raise 1.9ex\hbox{$...$}\kern -1.0em \hbox{$M$}}
\newcommand{\dddQ}{\kern 0.2em \raise 1.9ex\hbox{$...$}\kern -1.0em \hbox{$Q$}}
\newcommand{\dddI}{\kern 0.2em \raise 1.9ex\hbox{$...$}\kern -1.0em\hbox{$I$}}
\newcommand{\dddJ}{\kern 0.2em \raise 1.9ex\hbox{$...$}\kern-1.0em
\hbox{$J$}}
\newcommand{\dddcalJ}{\kern 0.2em \raise 1.9ex\hbox{$...$}\kern-1.0em
\hbox{${\cal J}$}}

\newcommand{\dddO}{\kern 0.2em \raise 1.9ex\hbox{$...$}\kern -1.0em
\hbox{${\cal O}$}}
\def\dddz{\raise 1.5ex\hbox{$...$}\kern -0.8em \hbox{$z$}}
\def\dddd{\raise 1.8ex\hbox{$...$}\kern -0.8em \hbox{$d$}}
\def\dddbd{\raise 1.8ex\hbox{$...$}\kern -0.8em \hbox{${\bf d}$}}
\def\ddbd{\raise 1.8ex\hbox{$..$}\kern -0.8em \hbox{${\bf d}$}}
\def\dddx{\raise 1.6ex\hbox{$...$}\kern -0.8em \hbox{$x$}}

\newcommand{\oma}{\Omega_{M}}
\newcommand{\ora}{\Omega_{R}}

\newcommand{\ola}{\Omega_{\Lambda}}

\newcommand{\rde}{\rho_{\rm DE}}
\newcommand{\wde}{w_{\rm DE}}

\graphicspath{ {./Graphs/} }

\begin{document}


\title{Probing modified gravitational wave propagation with \\strongly lensed coalescing binaries}



\author{Andreas Finke}
\author{Stefano Foffa}
\author{Francesco Iacovelli}
\author{Michele Maggiore}
\author{Michele Mancarella}

\affiliation{D\'epartement de Physique Th\'eorique and Center for Astroparticle Physics, Universit\'e de Gen\`eve, 24 quai Ansermet, CH--1211 Gen\`eve 4, Switzerland}



\begin{abstract}
It has been recently shown that quadruply lensed gravitational-wave (GW) events due to coalescing binaries can be localized to one or just a few galaxies, even in the absence of an electromagnetic counterpart. We discuss how this can be used to extract information on modified GW propagation, which is a crucial signature of   modifications of gravity at cosmological scales. We show that, using quadruply lensed systems, it is possible to constrain the parameter $\Xi_0$ that characterizes modified GW propagation, without the need of imposing a prior on $H_0$. A LIGO/Virgo/Kagra network at target sensitivity might already get a significant  measurement of $\Xi_0$, while  a third generation GW detector such as the Einstein Telescope could reach a very interesting accuracy.
\end{abstract}

\pacs{}

\maketitle

\section{Introduction}

Gravitational waves (GWs) propagating across cosmological distances are bent by nearby  galaxies and galaxy clusters,  just as light does. Signatures for lensing of GWs  have been searched in the data of the O1 and O2 LIGO/Virgo observing runs~\cite{Hannuksela:2019kle,Li:2019osa,McIsaac:2019use,Pang:2020qow,Liu:2020par,Dai:2020tpj} and in the O3a data~\cite{LIGOScientific:2021izm}, and  no compelling evidence for  lensing of GW events has   been reported. With current detector sensitivity, the expected fraction of observed strong lensing  events is of order $10^{-4}-10^{-3}$, depending on assumptions of the merger rate~\cite{LIGOScientific:2021izm} so, with the ${\cal O}(50)$ coalescences detected to date, the observation of a lensed event is unlikely (see  \cite{Diego:2021fyd} for an opposite viewpoint).
At design sensitivity, the rate of strongly lensed events for advanced LIGO/Virgo could be around one  event per year~\cite{Ng:2017yiu,Li:2018prc,Oguri:2018muv} (again, with a significant uncertainty due to the merger-rate density at high redshift). This figure greatly increases for  a third generation GW detector such as the Einstein Telescope (ET)~\cite{Punturo:2010zz,Maggiore:2019uih}, which could observe  ${\cal O}(50-150)$ strongly lensed binary black holes (BBHs) per year~\cite{Biesiada:2014kwa}, plus tens to hundreds BBH  events per year and 
${\cal O}(10-50)$ black hole-neutron star (BH-NS) and neutron star-neutron star (NS-NS) lensed events  per year,
that would have been below the detection threshold, but rise above the detection threshold  because of the lensing magnification~\cite{Ding:2015uha}. 

Just as for lensing of electromagnetic signals, which is now a standard tool in astrophysics~\cite{Treu:2010uj,Bartelmann:2010fz},  lensing of GWs could be used as a probe 
of several effects. In particular, the detection of multiply lensed GW events would provide a new tool for cosmography, allowing a measurement of the Hubble parameter $H_0$ and of the dark energy equation of state~\cite{Sereno:2011ty,Liao:2017ioi,Liu:2019dds,Li:2019rns}. In this paper we explore another potentially very interesting application of strong lensing of GW events, to the study of modified GW propagation. 

The paper is organized as follows. In sect.~\ref{sect:MGWp} we review the phenomenon of modified GW propagation, and the existing observational limits. Strong lensing of GWs is briefly reviewed in sect.~\ref{sect:sl}, focusing in particular on the potential for cosmography of quadruply lensed events, recently explored in~\cite{Hannuksela:2020xor}. In sect.~\ref{sect:SLinModGrav} we discuss strong lensing in modified gravity, pointing out that quadruply lensed events allow for the simultaneous determination of the standard (`electromagnetic') luminosity distance, and of the `GW luminosity distance' that characterizes modified GW propagation. In sect.~\ref{sect:forecasts} we discuss how, in practice,  one could use 
a quadruply lensed event to measure the  parameter $\Xi_0$ that we use to parametrize modified GW propagation. Sect.~\ref{sect:conclusions} contains our conclusions.

\section{Modified GW propagation} \label{sect:MGWp}

In recent years, modified GW propagation has been understood as a completely generic property of theories that modify   General Relativity (GR) on cosmological scales~\cite{Saltas:2014dha,Lombriser:2015sxa,Nishizawa:2017nef,Arai:2017hxj,Belgacem:2017ihm,Amendola:2017ovw,Belgacem:2018lbp,Belgacem:2019pkk}, and it has been realized that it can be the most promising observable for testing such theories (see, in particular, the discussions 
in~\cite{Belgacem:2017ihm,Belgacem:2018lbp,Finke:2021aom}).
 
The phenomenon is encoded in a modification of the `friction term' in the equation that governs
the propagation of tensor perturbations over a Friedmann-Robertson-Walker (FRW) background, that, in modified gravity, can take the form 
\be\label{prophmodgrav}
\tilde{h}''_A  +2 {\cal H}[1-\delta(\eta)] \tilde{h}'_A+k^2c^2\tilde{h}_A=0\, .
\ee
GR is recovered for $\delta(\eta)=0$ (we use standard notation: $h_A$ is the GW amplitude, $A=+,\times$ labels the two polarizations, the prime denotes the derivative with respect to cosmic time $\eta$,  $a(\eta)$ is the FRW scale factor, and  ${\cal H}=a'/a$). We have not included the possibility of modifying the coefficient  of the term
$k^2c^2$ term, since this would affect  the speed of GWs, and is now excluded, 
at a level  $O(10^{-15})$   \cite{Monitor:2017mdv}. 
Still,  even in modified gravity theories  that do not change the $k^2c^2$ term (such as a subset of scalar-tensor theories~\cite{Creminelli:2017sry,Sakstein:2017xjx,Ezquiaga:2017ekz,Baker:2017hug}, or modifications of GR induced by the generation of non-local terms in the quantum effective 
action~\cite{Maggiore:2013mea,Maggiore:2016gpx,Belgacem:2020pdz}), a non-trivial function $\delta(\eta)$ emerges. Indeed, 
the detailed study in \cite{Belgacem:2019pkk} shows that this modification  is completely generic,  and  takes place in all  modified gravity models that have been investigated.

A consequence of \eq{prophmodgrav}, with a non-vanishing function $\delta(\eta)$, is that, in  the propagation  across cosmological distances, the amplitude of GWs is attenuated in a different way, with respect to the $\propto 1/a$ behavior of GR. Then, 
 GW observations of coalescing binaries no longer measure  the  luminosity distance $d_L(z)$ of the source [that, in this context, we  call the `electromagnetic luminosity distance' and we denote by $\dem(z)$], but rather a  `GW luminosity distance'  $\dgw(z)$, related to $\dem(z)$ by~\cite{Belgacem:2017ihm,Belgacem:2018lbp}
\be\label{dLgwdLem}
\dgw(z)=\dem(z)\exp\left\{-\int_0^z \,\frac{dz'}{1+z'}\,\delta(z')\right\}\, ,
\ee
where the function $\delta$ that appears in \eq{prophmodgrav} has now been written as a function of redshift. 

For  comparing with observations  it is useful to have a parametrization of this effect in terms of a small number of parameters, rather than a full function $\delta(z)$. In the following we will use the parametrization proposed in~\cite{Belgacem:2018lbp},
\be\label{eq:fit}
\frac{d_L^{\,\rm gw}(z)}{d_L^{\,\rm em}(z)} \equiv\Xi(z)=\Xi_0 +\frac{1-\Xi_0}{(1+z)^n} \, ,
\ee
in terms of two parameters $(\Xi_0,n)$.
This parametrization reproduces the  fact that, for $z=0$, $d_L^{\,\rm gw}/d_L^{\,\rm em}= 1$, as it should,  and 
is such that,
in the limit of large redshifts, $d_L^{\,\rm gw}/d_L^{\,\rm em}$ goes to a constant value $\Xi_0$. This is   indeed what happens in most modified gravity models,  where  the deviations from GR only appear in the recent cosmological epoch; therefore, $\delta(z)$ goes to zero at large redshifts, and the integral in \eq{dLgwdLem} saturates to a constant value. \Eq{eq:fit} smoothly interpolates between these two limiting behaviors with a power-law that, written in terms of the scale factor $a$, reads, even more simply,
$d_L^{\,\rm gw}/d_L^{\,\rm em}=\Xi_0+a^n (1-\Xi_0)$.
As shown in ref.~\cite{Belgacem:2019pkk}, \eq{eq:fit} fits remarkably well the explicit results from all the best-studied modified gravity models, such as several examples of
Horndeski and DHOST theories, or non-local infrared modifications of gravity [with the exception of bigravity, where the result for  $d_L^{\,\rm gw}(z)/d_L^{\,\rm em}(z)$ displays some oscillations as a function of redshift, rather than the smooth behavior given by \eq{eq:fit}]. In the parametrization 
(\ref{eq:fit}), GR is recovered 
when $\Xi_0=1$ (for all $n$).

The next question is  what is the  size of the deviations  from the GR value $\Xi_0=1$, that one could expect in viable modified gravity models. In general, to comply with cosmological observations with  electromagnetic probes such as 
cosmic microwave background (CMB), supernovae (SNe), baryon acoustic oscillations (BAO), and structure formation, the deviations from GR and, more specifically, from $\Lambda$CDM,  are bounded at the level of  a few percent, for both the background evolution and    the scalar perturbations. One might then expect that also the  deviations for tensor perturbations (i.e., for GWs propagating in a FRW background)  will be of the same order,  leading at most to a deviation of $\Xi_0$ from 1 by a few percent. 
However, the study of explicit models shows that this is not necessarily the case, and the deviations in the tensor sector can be much larger. A remarkable example is given  by the so-called RT non-local gravity model  \cite{Maggiore:2013mea} (see \cite{Belgacem:2020pdz} for an updated review). This model is very close to $\Lambda$CDM in the background and in the scalar perturbations, and indeed fits cosmological observations at a level statistically equivalent to $\Lambda$CDM. However, it predicts that
$\Xi_0$ can be as large as $1.80$
\cite{Belgacem:2019lwx,Belgacem:2020pdz}, corresponding to a $80\%$ deviation from the GR value $\Xi_0=1$. Independently of the specific virtues of this model (which has an appealing field-theoretical motivation, with non-local terms generated by infrared quantum effects in the quantum effective action, generates dynamically a dark energy, and fits current observations at the same level as $\Lambda$CDM), this can be taken just as an example of the fact that phenomenologically viable models can predict large deviations from $\Lambda$CDM in the sector of tensor perturbations, which is the new cosmological window that GW experiments are beginning to open.

An observational limit on modified GW propagation was obtained in~\cite{Belgacem:2018lbp} from 
GW170817,   comparing
the GW luminosity distance, obtained from the GW observation, to the electromagnetic luminosity distance of the galaxy hosting the counterpart,  obtained   from surface brightness fluctuations. Given the small redshift of GW170817, this is really a measurement of $\delta(z=0)$, independently of any parametrization, and gives
$\delta(0)=-7.8^{+9.7}_{-18.4}$, which is of course consistent with the GR prediction $\delta(z)=0$. 
Setting  a value  $n\simeq 2$ (typical for instance of the predictions of nonlocal gravity) and using $\delta(0)=n(1-\Xi_0)$, this can be translated into a bound $\Xi_0\,\lsim\, 14$ (see also \cite{Arai:2017hxj,Lagos:2019kds} for other analysis using GW170817, resulting in somewhat broader but consistent limits).

Recently, a more stringent limit has been obtained in \cite{Finke:2021aom}, using  BBH coalescences from the O1, O2 and O3a LIGO/Virgo run and correlating them with the GLADE galaxy catalog~\cite{Dalya:2018cnd}. The result is $\Xi_0=2.1^{+3.2}_{-1.2}$ ($68\%$ c.l.). This shows that interesting limits on $\Xi_0$ can already be obtained with current GW observations. An even more stringent result is obtained under the tentative  identification of the flare  ZTF19abanrhr as the electromagnetic counterpart of the BBH coalescence GW190521, in which case the result obtained in \cite{Finke:2021aom} is  $\Xi_0=1.8^{+0.9}_{-0.6}$ (see also \cite{Mastrogiovanni:2020mvm} for a related result, and  \cite{Mastrogiovanni:2020gua} for forecasts on the limits that could be obtained with ${\cal O}(100)$ BNS).  Modified GW propagation has also been recently constrained by combining GW data with the mass distribution of compact binaries \cite{Ezquiaga:2021ayr}, following the strategy proposed in \cite{Farr:2019twy}.

For third-generation detectors such as  the  Einstein Telescope (ET)~\cite{Punturo:2010zz,Maggiore:2019uih} and Cosmic Explorer~\cite{Reitze:2019iox}, or for  the LISA space interferometer~\cite{Audley:2017drz}, the perspectives are quite exciting. In  \cite{Belgacem:2018lbp,Belgacem:2019tbw}, using 
 state-of-the-art mock catalogs for the GW events and for the detection of an associated GRB, it is estimated that  ET could measure $\Xi_0$ to  about $1\%$ (depending on the network of GRB satellites available at the time when ET will operate). Forecasts for LISA have been presented 
in \cite{Belgacem:2019pkk}, using  the coalescence of supermassive  black holes (SMBH),  and it has been found that $\Xi_0$  could be measured  to 
$(1-4)\%$ accuracy. Forecasts for the correlation between LISA SMBH events and the prediction of the Horndeski theory for large scale structure have been presented in \cite{Baker:2020apq}.

In the following we will discuss another possible way of searching for modified GW propagation, using strongly lensed GW events. 
%
\section{Strong lensing of GW events}\label{sect:sl}

In GW astronomy the angular resolutions of the detectors are  orders of magnitude above the arcsec resolution   needed to spatially separate typical lensed  images. However, the signals from coalescing binaries have a short duration. Therefore,
strong lensing will rather manifest itself  through repeated GW detections, associated to signals reaching the observer through different paths from the same source. These events  will have a relative time delay,  of orders of minutes to months for lensing by galaxies and up to years for lensing by galaxy clusters,  and a  different  amplitude, which reflects the fact that  signals traveling through different paths  undergo different amount of magnification or demagnification. The other parameters entering the  waveform, such as the detector frame masses, spins, sky locations, etc., are the same (except, possibly, for a frequency-independent phase shift), and a Bayesian analysis can be performed  to compare the hypothesis that two (or more) GW events belong to the same source with the hypothesis that they are unrelated~\cite{Hannuksela:2019kle,Li:2019osa,McIsaac:2019use,Pang:2020qow,Liu:2020par,Dai:2020tpj,Pagano:2020rwj,Smith:2017mqu,Haris:2018vmn}. 

The possibility of using strongly lensed GW events with an electromagnetic counterpart to perform cosmography has been discussed in \cite{Liao:2017ioi,Liu:2019dds}, but BBH coalescences are not expected, in general, to have electromagnetic counterparts. However,
as shown in \cite{Hannuksela:2020xor}, quadruply lensed events (which are estimated to be about 30\% of the total lensed events at LIGO/Virgo, and 6\% for ET~\cite{Li:2018prc}) are especially interesting, because they can be used as standard sirens even in the absence of an associated electromagnetic flare, by comparing the lensing pattern obtained from GW observations with that  observed electromagnetically. 
The reason is that quadruply lensed  systems have three independent time delays and three independent magnification ratios, and this provides sufficient information to reconstruct the source-lens system from the GW observations. The corresponding lensing pattern must then show up in the electromagnetic observation
of the source-lens system.
Assuming  that the GW event originates from a galaxy that emits electromagnetic radiation, and that 
(possibly with dedicated follow-up) we know from electromagnetic observations all the strong lensing systems in the localization area of the GW event, taken to be given by 
the localization resolution of a LIGO/Virgo/KAGRA network at target sensitivity, ref.~\cite{Hannuksela:2020xor} shows that it is possible to localize the GW source to a single, or at most a few, galaxies. Spectroscopic follow-up could  then provide an accurate determination of  the redshifts $z_l$ of the lens and $z_s$ of the source. Once identified the GW host galaxy, a detailed lens model can be used to de-lens the system and infer the absolute magnifications $\mu_i$ associated with the  distinct images.

According to gravitational lensing theory (see e.g. \cite{1992grle.book.....S,Treu:2010uj,Bartelmann:2010fz}), the time delay between two `images' $i$ and $j$ is given by
\be
c\Delta t_{i,j}=(1+z_l)D_{\Delta t}(z_l,z_s)\Delta\phi_{i,j}\, ,
\ee
where $\Delta\phi_{i,j}$ is the difference of the Fermat potentials between the $i$ and $j$ images, which is obtained from the lens reconstruction, and
\be\label{DDeltat}
D_{\Delta t}(z_l,z_s)=\frac{d_A(z_l)d_A(z_s)}{d_A(z_l,z_s)}\, ,
\ee
where $d_A(z_l)$, $d_A(z_s)$ and $d_A(z_l,z_s)$ are the angular diameter distances from the lens to the observer, from the source to the observer, and from the lens to the source, respectively. 
Furthermore, the observed luminosity distance extracted from the $i$-th lensed image, $D_i$, is related to the actual, unlensed, luminosity distance $d_L$ of the source by 
\be\label{DiGW}
D_i=d_L/\sqrt{\mu_i}\, .
\ee 
Since the ratio of time delays, and the ratio of magnifications $\mu_i/\mu_j$, are sufficient to reconstruct the source position, the absolute scale of the time delays and of the absolute magnifications can  be used to perform inference on the cosmology.  This strategy has been used in ~\cite{Hannuksela:2020xor} to forecast the accuracy that could be obtained on $H_0$, in the context of $\Lambda$CDM,  with a LIGO/Virgo/KAGRA network.

%
\section{Strong lensing in modified gravity} \label{sect:SLinModGrav}

In this paper we  examine the information that can be obtained  on $\Xi_0$, in the context of modified gravity, from a quadruply lensed GW event. The crucial observation is that, in modified gravity, 
\eq{DiGW} becomes  
\be\label{Didgw}
D_i=\dgw/\sqrt{\mu_i}\, ,
\ee 
since $\dgw$ is the quantity that encodes the attenuation of the GW amplitude in the propagation across cosmological distances. In contrast, the angular diameter distances that appear in \eq{DDeltat} are the same for GWs and for electromagnetic waves, and have nothing to do with the damping of the GW amplitude during their propagation.
These angular diameter distances  are therefore related to the standard (`electromagnetic') luminosity distance by the usual relation $d_A(z)=\dem(z)/(1+z)^2$. In particular, recalling that (in a spatially flat universe)
\be
d_A(z_l,z_s)=d_A(z_s)-\frac{1+z_l}{1+z_s} d_A(z_l)\, ,
\ee
we can rewrite $D_{\Delta t}(z_l,z_s)$ in terms of the electromagnetic luminosity distances of the source and of the lens, as  
\be\label{DDeltadem}
D_{\Delta t}(z_l,z_s)=\frac{\dem(z_l)\dem(z_s)}{(1+z_l)^2\dem(z_s)-(1+z_l)(1+z_s)\dem(z_l)}
\, ,
\ee
(see also~\cite{Cao:2020sky}, where a study of the viscosity term was carried out for GW events with counterpart). We write 
\be\label{eq:dem}
\dem(z)=\frac{1+z}{H_0}\int_0^z\, 
\frac{d\tilde{z}}{E(\tilde{z})}\, ,
\ee
where
\be
E(z)=[\ora (1+z)^4+\oma (1+z)^3+\rde(z)/\rho_0 ]^{1/2}\, ,
\ee
$\rho_0=3H_0^2/(8\pi G)$, $\ora$ and $\oma$ are the radiation and matter density fractions, respectively, while $\rde(z)$ is the dark energy (DE) density. In $\Lambda$CDM $\rde(z)/\rho_0=\ola$ is a constant, while in a generic modified gravity theory
\be\label{rhoDE}
\rde(z) /\rho_0 =\Omega_{\rm DE}\, e^{ 3\int_{0}^z\, \frac{d\tilde{z}}{1+\tilde{z}}\, [1+\wde(\tilde{z})]}\, ,
\ee 
where $\Omega_{\rm DE}=\rde(0)/\rho_0$, and $\wde(z)$ the DE equation of state.
Using \eq{eq:dem}, \eq{DDeltadem} can be rewritten as 
\be\label{DDtdem}
D_{\Delta t}(z_l,z_s)=\frac{R(z_l,z_s)}{(1+z_l)(1+z_s)}\, \dem(z_s)\, ,
\ee
where
\be
R(z_l,z_s)=\frac{\int_0^{z_l} dz'/E(z')}{\int_{z_l}^{z_s} dz'/E(z')}\, .
\ee
Observe that $H_0$ canceled in the ratio $R(z_l,z_s)$ which therefore depends on the cosmology only through $\oma$  and, in modified gravity, on  the DE equation of state, that enters in $E(z)$ through $\rde(z)$.

To sum up, from a quadruply lensed GW signal, we can get a reconstruction of the redshift of the source and of the lens, four  measurements of $\dgw$ from the amplitude of the four images,  and a measurement of the combination (\ref{DDeltadem}), that (given $z_l$, $z_s$ and the cosmology) only involves the electromagnetic luminosity distance of the source. The DE sector of the theory enters through the function $\wde(z)$ that affects $\dem(z)$ according to \eqst{eq:dem}{rhoDE}, and through the function $\Xi(z)$ that, given $\dem(z)$, determines $\dgw(z)$ according to \eq{eq:fit}. Therefore, given the (four) observed values of $\dgw$ and the observed value of $D_{\Delta t}(z_l,z_s)$, we can in principle perform an inference on  $\wde(z)$, using for instance the $(w_0,w_a)$ parametrization  of the redshift dependence \cite{Chevallier:2000qy,Linder:2002et},
$w_{\rm DE}(z)= w_0+[z/(1+z)] w_a$, and on $\Xi(z)$, for which we will use the $(\Xi_0,n)$ parametrization (\ref{eq:fit}).

In the following, to simplify the analysis, we will fix $\wde=-1$ (i.e. $w_0=-1, w_a=0$), as in $\Lambda$CDM, and we will only consider $\Xi_0$ as a free parameter.\footnote{The parameter $n$ in \eq{eq:fit} describes the precise power-law behavior between the asymptotic regimes at small and large $z$ and its precise value is less important. When producing the plots in sect.~\ref{sect:forecasts}, we will fix it for definiteness to the value $n=1.91$,which is the value predicted by the RT model in the same limit in which $\Xi_0=1.80$.}
This is motivated by the fact that we know, from cosmological observations with electromagnetic probes such as  cosmic microwave background (CMB), supernovae (SNe), baryon acoustic oscillations (BAO), structure formation, etc., that $\wde$ cannot differ from $-1$ by more than about $5\%$. In contrast, the current limit on $\Xi_0$, that can come only from GW observations, are much broader. As mentioned in sect.~\ref{sect:MGWp}, the correlation of O1+O2+O3a dark sirens with the GLADE galaxy catalog gives $\Xi_0=2.1^{+3.2}_{-1.2}$, which still allows (at $1\sigma$) values of $\Xi_0$ as large as $\Xi_0\simeq 5$, i.e. deviations from the GR value at the level of  $500\%$. The most stringent values obtained in \cite{Mastrogiovanni:2020mvm,Finke:2021aom} are based on the assumption  that the flare  ZTF19abanrhr is the  counterpart of the BBH coalescence GW190521, an identification which is not secure, and in any case still allow deviations at the 
$300\%$ level from GR. As we have mentioned above,  a phenomenologically viable model such as the RT nonlocal model predicts a value of $\Xi_0$ that can be as large as $1.80$, i.e. a $80\%$ deviation from GR. 

Furthermore, as discussed in \cite{Belgacem:2018lbp}, the accuracy that can be reached on $\Xi_0$ from standard sirens is in general better than the corresponding accuracy on $w_0$. This is due to  to the fact that the effect of $w_0$ on the electromagnetic luminosity distance is masked  by partial degeneracies with $H_0$ and $\oma$: moving $w_0$ away from $-1$, the value of $H_0$ and $\oma$ (which are determined by fitting the modified  model to cosmological data
such as CMB, BAO and SNe) change, with respect to their values in $\Lambda$CDM, and precisely in the direction to partially compensate the effect of the change of $w_0$ in the luminosity distance.

 For these reasons, it makes sense to consider a phenomenological scenario where the deviation of $\Xi_0$ from 1 are large, while $\wde$ is very close to $-1$ (and the RT nonlocal  model provides an explicit realization of this scenario). Note that, in contrast, when studying cosmography with GWs, it is not really meaningful to  consider $w_0$ as a free parameter, while neglecting modified GW propagation. Whenever a dynamical DE model predicts $w_0\neq -1$, it also predicts $\Xi_0\neq 1$, and the latter effect in general dominates, possibly by even one or two orders of magnitude.\footnote{In \cite{Dalang:2019fma} it has been suggested that, in models that require screening (which does not include the RT nonlocal gravity model) the screening mechanism could also eliminate any detectable modification of the luminosity distance. This, however, applies only in theories where the  Newton constant that describes the coupling of matter is the same as the effective Newton constant that appears in the quadratic self-interaction term of gravitational waves, which is not the case in generic modified gravity theories~\cite{Baker:2020apq}.}

\section{Extracting $\Xi_0$ from quadruply lensed events}\label{sect:forecasts}

We now discuss how, in practice, modified GW propagation could be tested from a quadruply lensed system. The first obvious possibility is to use the redshift $z_s$ of the source, obtained from the reconstruction of the source-lens system and possibly a dedicated spectroscopic follow-up (in which case the error on $z_s$ would be negligible with respect to the other uncertainties), {\em assume} a value of $H_0$ (and of $\oma$), and infer from this the electromagnetic luminosity distance to the source, $\dem(z_s)$. We can then compare it with the GW measurement of $\dgw(z_s)$, obtained combining the  four measurements of the amplitudes $D_i$ in \eq{Didgw}, with the $\mu_i$ given by the reconstruction of the lens system. 
The posteriors for $\dgw(z_s)$ and $\dem(z_s)$ could then be transformed into a posterior  $P[(\Xi(z_s)]$  for $\Xi(z_s)$ and therefore  (assuming a given $n$) into a posterior for $\Xi_0$, 
obtained from 
\bees
P(\Xi_0)&=&P[(\Xi(z_s)]\, \frac{d\,\Xi(z_s)}{d\,\Xi_0}\nn\\
&=&P[(\Xi(z_s)] \[ 1-\frac{1}{(1+z_s)^n}\]\, .\label{PostXi0}
\ees
More generally, we can express the result in terms of  a posterior in the $(\Xi_0,n)$ plane.

The disadvantage of this procedure is that we need to assume a value for $H_0$. Even neglecting for a moment the tension between early-Universe~\cite{Aghanim:2018eyx,Abbott:2018xao}
and late Universe~\cite{Riess:2019cxk,Wong:2019kwg,Riess:2020fzl} measurements of $H_0$, and focusing on early Universe measurement only, still the value of $H_0$ (and of $\oma$), determined by fitting the cosmological parameters  to  CMB, BAO and SNa data, depends in principle on the modified gravity theory considered. This might not be such a problem, at least as long as we look for large deviations of $\Xi_0$ from the GR value, since typical  modified gravity models (or, at least, those based on late-time DE) that fit current cosmological data predict a value of $H_0$ quite close to that of $\Lambda$CDM. For instance, in the RT nonlocal model, the mean values of $H_0$ and $\oma$ obtained from a fit to CMB, BAO and SNa data differ from that obtained in  $\Lambda$CDM  by less than $0.1\%$. Nevertheless, in view of the discrepancy between early- and late-Universe measurements of $H_0$, it would be even more interesting to have a determination of $\Xi_0$ that is independent on assumptions on $H_0$.

This is indeed possible, thanks to the extra information carried by  $D_{\Delta t}(z_l,z_s)$: according to \eq{DDtdem}, after lens reconstruction, a measurement of $D_{\Delta t}(z_l,z_s)$
determines directly $\dem(z_s)$
(assuming that  $\wde$ is sufficiently close to $\wde=-1$, so that we can  simply set $\wde=-1$, at least as long as we search for values of $\Xi_0$ significantly different from the GR value $\Xi_0=1$; otherwise, we will have to perform a joint inference on $\Xi_0$ and $w_0$).

Of course, the accuracy of the measurement  that one can obtain for $\dgw$ and  for $D_{\Delta t}(z_l,z_s)$ strongly depends  on the network of detectors under consideration, and  will also in general  change significantly from event to event. As an illustration of the procedure, we first consider an accuracy on these measurements of the order of  that obtained from the detailed analysis in \cite{Hannuksela:2020xor} (see in particular their Fig.~5),  which assumes a LIGO/Virgo/Kagra network at target sensitivity,  simulates a lens distribution that follow a given galaxy-galaxy lens distribution~\cite{Collett:2015roa}, randomly simulates GW events that are quadruply lensed and pass a network detector threshold $\rho_{\rm thre}>10$, and performs the lens reconstruction.

\begin{figure}[t]
\centering
\includegraphics[width=0.5\textwidth]{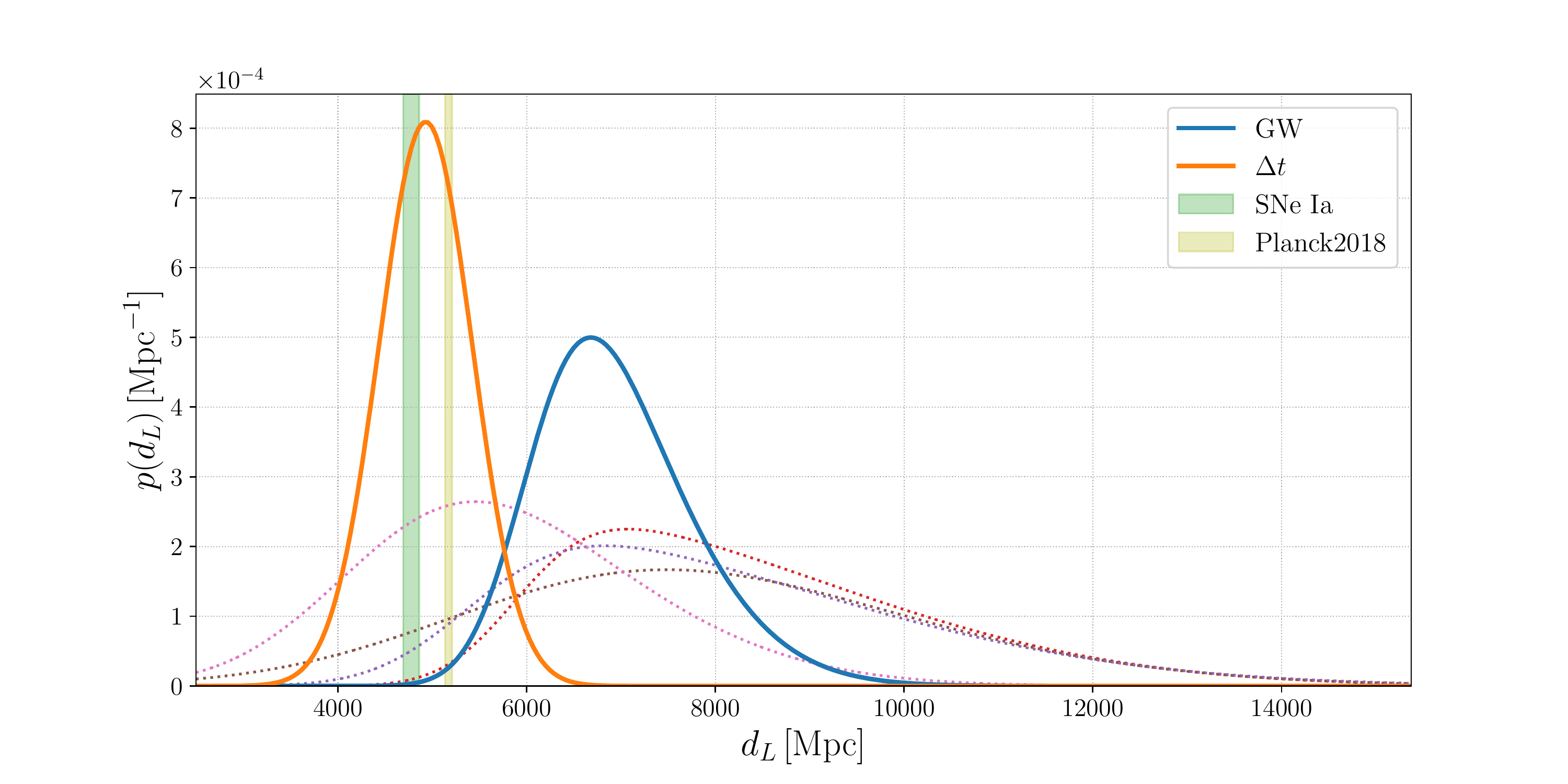}
\includegraphics[width=0.5\textwidth]{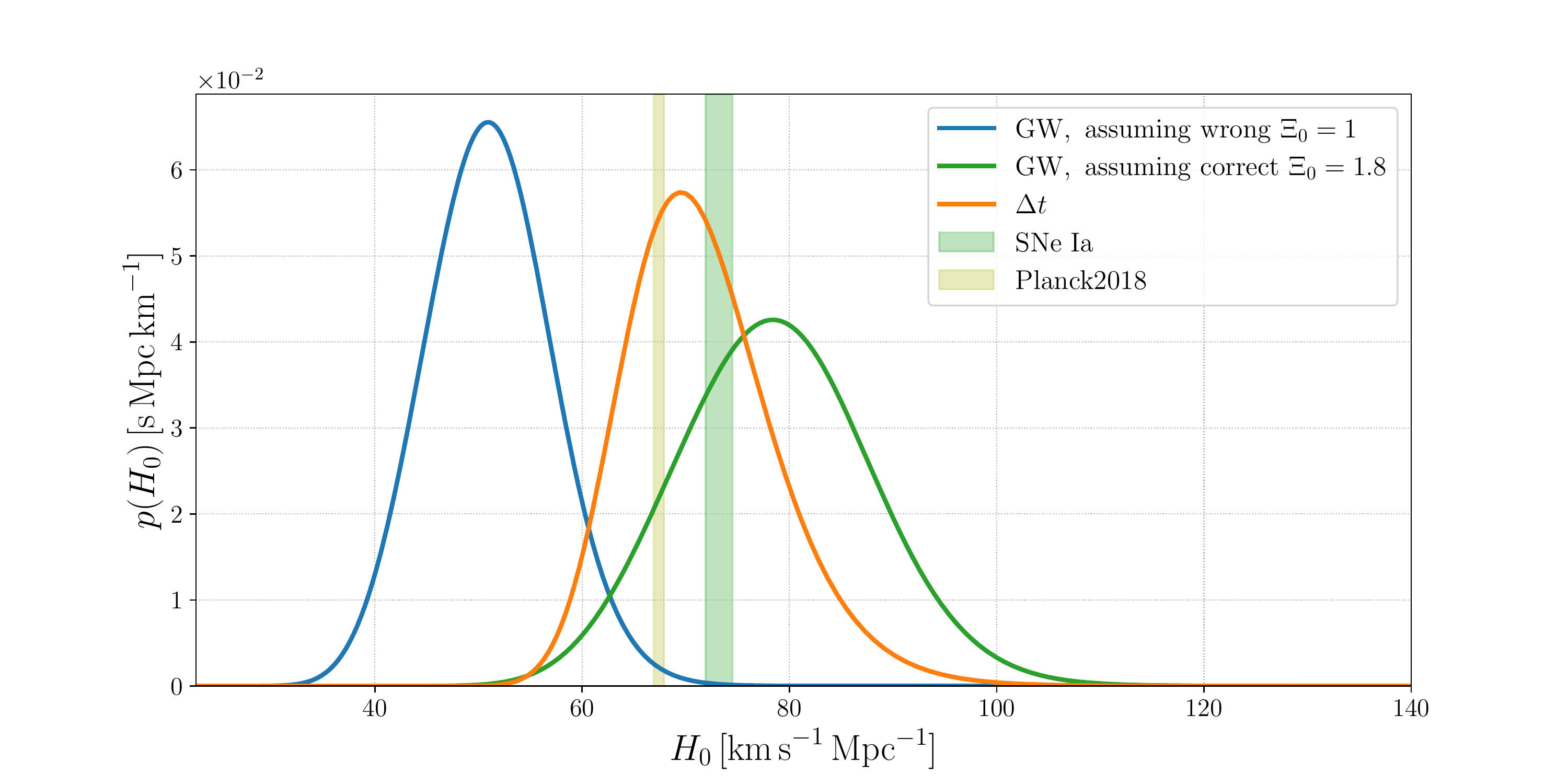}
\caption{Upper plot: posterior for $d_L$ as the result of combining a quadruple lensed GW (solid blue, each individual image represented by the dotted curves), and the measurement of time delays (orange). Lower plot: the same information translated into a posterior for $H_0$, assuming $\Lambda$CDM (blue) or a modified gravity model (green).}
\label{fig:dLH0}
\end{figure}

The upper panel of Fig.~\ref{fig:dLH0} shows the kind of posteriors that one could expect. To produce this plot we have assumed that the fiducial  values of the cosmological parameters  are $H_0=70\, {\rm km}\, {\rm s}^{-1}\, {\rm Mpc}^{-1}$, $\oma=0.31$ and $\Xi_0=1.8$. We are therefore assuming that the correct theory of gravity is a modified gravity theory with a large deviation from $\Xi_0=1$, inspired by the value found in the RT nonlocal model. We also assume that 
the `true' redshift of the source is $z_s=0.8$. We have then computed   the  `true' values of $\dem(z_s)$ and $\dgw(z_s)$ with our fiducial values of the cosmological parameters.
The `observed' mean value $\mu_{\rm em}$ of the luminosity distance $\dem$, obtained from 
$D_{\Delta t}(z_l,z_s)$, has been drawn  from a gaussian with mean $\dem(z_s)$ and standard deviation  
$\sigma =0.1\dem(z_s)$, to simulate a scattering of the observed value with respect to the true value. For the posterior of $\dem$, shown as the orange line in the upper panel of Fig.~\ref{fig:dLH0}, we have assumed again a gaussian shape,
as suggested by the result in \cite{Hannuksela:2020xor}, centered on $\mu_{\rm em}$ and with the same  width  $\sigma =0.1\dem(z_s)$, i.e. a measurement of the distance at the $10\%$ level. The green and light green bands show the result that would be obtained for $\dem(z_s)$ using the value  $H_0=67.4\pm 0.5\, {\rm km}\, {\rm s}^{-1}\, {\rm Mpc}^{-1}$ from {\em Planck \,} 2018~\cite{Aghanim:2018eyx} and the value $H_0=73.2\pm 1.3\, {\rm km}\, {\rm s}^{-1}\, {\rm Mpc}^{-1}$ from SNe~Ia~\cite{Riess:2020fzl}, respectively. Our fiducial value is $H_0=70\, {\rm km}\, {\rm s}^{-1}\, {\rm Mpc}^{-1}$, somewhat in the middle of them, and the random scattering that we have added happens to move the orange posterior closer to  the SNe~Ia value.

For  $\dgw$, the distribution of the individual measurements from each of the four images (dashed lines in the  upper panel  of Fig.~\ref{fig:dLH0}) are now taken to be  skewed gaussians, again  as suggested by the results in Fig.~5 of \cite{Hannuksela:2020xor}. For the mean values $(\mu_{\rm gw})_i$, we proceed as  for  $\dem$. The standard deviation and skewness of  the dashed lines are each drawn from a uniform distributions (between $40\%$ and $60\%$ relative error for the standard deviation, and between 0.5 and 8 for the skewness). The blue solid line is then the combination of these four distributions. Even if the separate distributions are skewed gaussians, the combined posterior for the four events is already fairly gaussian. The mean value for $\mu_{\rm gw}$ of the combined distribution is of course sensibly different from $\mu_{\rm em}$, since we have taken $\Xi_0=1.8$, and the quantity that is measured by the GW amplitude is a GW luminosity distance, not an electromagnetic luminosity distance.

Of course, when analyzing the data from a quadruply lensed system, one would first start by assuming the validity of GR, and therefore one would treat both the measurement from $D_{\Delta t}(z_l,z_s)$ and the measurement from the four separate amplitudes as measurements of the same quantity $d_L(z_s)$. The signal of a deviation from GR would then be given by the observation that the combined posterior distribution from the four separate amplitudes, blue line in the upper panel of Fig.~\ref{fig:dLH0}, is not consistent, to a given statistical significance, with the posterior obtained from $D_{\Delta t}(z_l,z_s)$, orange line  in the upper panel of Fig.~\ref{fig:dLH0}. 
In the lower panel of Fig.~\ref{fig:dLH0} we show the corresponding posteriors of $H_0$ that would be obtained from the posteriors of $d_L(z_s)$, using the fact that $z_s$ has been determined by the lensing reconstruction (assuming negligible error on $z_s$, thanks to a spectroscopic follow up), and assuming the validity of GR and (spatially flat)  $\Lambda$CDM, i.e. using
\eq{eq:dem} with $\rde(z)/\rho_0=\ola$ and our fiducial value of $\oma$. In this case, the signal of deviation from GR and   $\Lambda$CDM would be given by the  discrepancy between the blue curve, obtained  from the GW amplitudes assuming $\Xi_0=1$, and the orange curve,  obtained from $D_{\Delta t}(z_l,z_s)$. The green curve is the result from the measurement of the four GW amplitudes, once one uses our fiducial value $\Xi_0=1.8$, and we see that the consistency with the posterior from  $D_{\Delta t}(z_l,z_s)$ is re-established (within a small difference  that reflects the scatter in $\mu_{\rm gw}$ and $\mu_{\rm em}$ that we have introduced to simulate an actual observational situation).

\begin{figure}[t]
\centering
\includegraphics[width=0.5\textwidth]{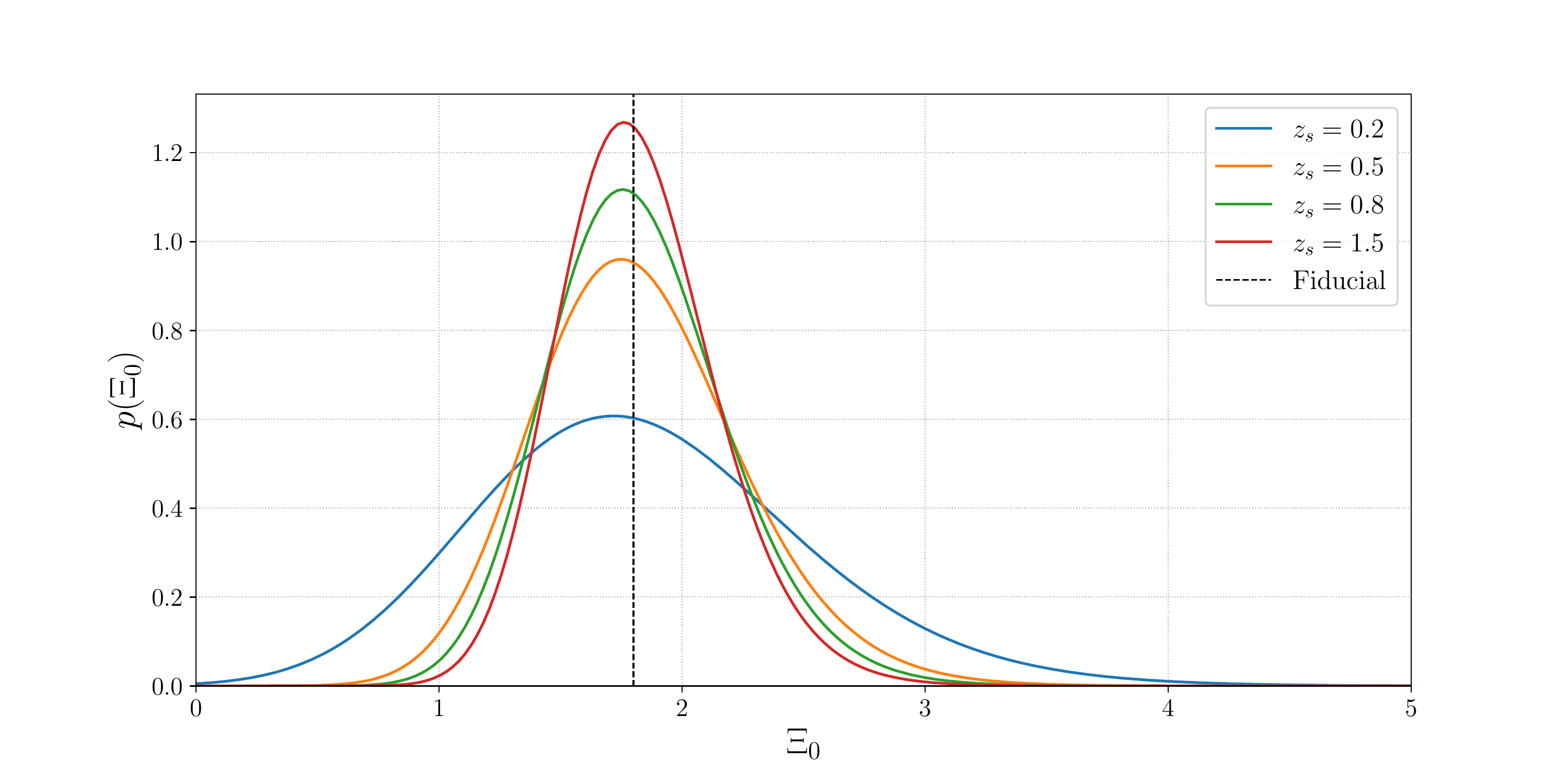}
\caption{Examples of posteriors for $\Xi_0$ for different values of the source redshift $z_s$,
corresponding to different possible measurements for a fiducial model with $\Xi_0=1.8$.}
\label{fig:zspan}
\end{figure}

After being alerted on the possibility of deviations from GR by a certain level of  inconsistency between these posteriors, one would then naturally open up the parameter space to include $\Xi_0$, interpreting the combined measurement from the four amplitudes as a determination of $\dgw(z_s)$, and the measurement from $D_{\Delta t}(z_l,z_s)$ as a determination of $\dem(z_s)$. From
this, one would then get  a posterior for $\Xi(z_s)=\dgw(z_s)/\dem(z_s)$, and then for $\Xi_0$, as in \eq{PostXi0}. The corresponding result is shown in Fig.~\ref{fig:zspan}, where we show the posterior obtained from the simulated source at $z_s=0.8$ already  used in Fig.~\ref{fig:dLH0} (except that here we do not add a random displacement, so that the median of the distribution is now centered exactly on our fiducial value $\Xi_0=1.8$), as well as the posteriors obtained with the same procedure for sources at different values of $z_s$.  For all four  redshifts in the plot, we have assumed the same fractional error on the measurements of
$D_{\Delta t}(z_l,z_s)$ and on the GW luminosity distance, as in the case $z_s=0.8$.  We see that, at fixed relative error in the measurement, the posterior for $\Xi_0$ becomes more and more narrow increasing $z_s$, since the ratio $\dgw(z)/\dem(z)$ increases with $z$, at least before saturating at the asymptotic value $\Xi_0$. Observe also that, since we did not add any random displacement, 
the medians of the posteriors  in Fig.~\ref{fig:zspan}
coincide with the  fiducial value of $\Xi_0$. In reality, instrument noise and uncertainties in the reconstruction will displace both the posteriors of $\dem(z_s)$ and of $\dgw(z_s)$, and half of the events (when the peaks are moved apart from each other by chance) will therefore provide more tension with the GR value  $\Xi_0=1$ than what we illustrate, while the other half will provide less tension.

\begin{figure}[t]
\centering
\includegraphics[width=0.5\textwidth]{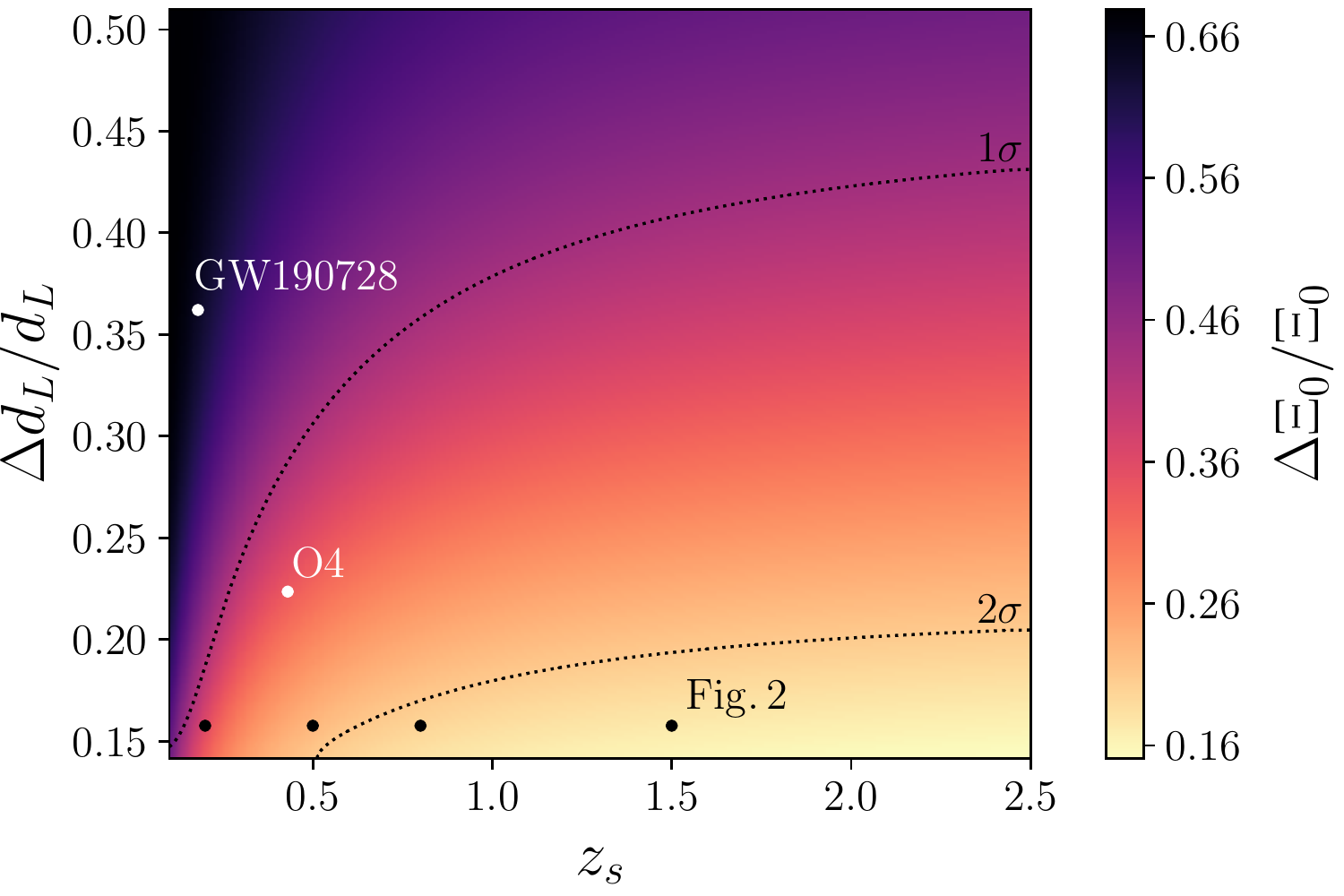}
\caption{Colormap plot of the precision attainable on $\Xi_0$ as a function of the redshift and of the combined relative error of the measurements coming from the GW amplitude and arrival times, assuming $\Xi_0=1.8$ as fiducial value. See the text for the meaning of the different dots and lines.} 
\label{fig:mesh}
\end{figure}

Fig.~\ref{fig:zspan} has been made just as an illustration of the procedure for  sources at a few given redshifts and  given observational errors  on the the electromagnetic and gravitational luminosity distances. To have a more general understanding,  we repeat the analysis on a grid of values of source redshifts $z_s$ and different values for the relative errors on the measurement of the GW luminosity distance from the combined four GW amplitudes, $(\Delta d_L/d_L)_{\rm GW}$,  and of the error on the electromagnetic luminosity distance from $D_{\Delta t}(z_l,z_s)$, $(\Delta d_L/d_L)_{{\Delta t}}$. Actually, in a first approximation, the result depends mainly on the sum of the relative errors in quadrature,
\be\label{Deltatsudtot}
\frac{\Delta d_L}{d_L}\equiv \[  \(\frac{\Delta d_L}{d_L} \)_{\rm GW}^2
+  \(\frac{\Delta d_L}{d_L} \)_{\Delta t}^2 \]^{1/2}\, .
\ee
In the following, we  fix $(\Delta d_L/d_L)_{\Delta t}=10\%$ and we vary 
$(\Delta d_L/d_L)_{\rm GW}$, but the result applies to any other combinations of 
$(\Delta d_L/d_L)_{\Delta t}$ and  
$(\Delta d_L/d_L)_{\rm GW}$ that gives the same total $\Delta d_L/d_L$.

In  Fig.~\ref{fig:mesh} we show the result for the relative error $\Delta \Xi_0/\Xi_0$ in the plane $(z_s, \Delta d_L/d_L)$, where,
for each point, $\Delta\Xi_0$ is the $68\%$ confidence interval  of the distribution of $\Xi_0$, and $\Xi_0$ is the median of the distribution, and we use a range of redshifts appropriate to 2G detectors (note that lensing magnification can somewhat extend their range). In the figure,
the black dots correspond to the examples shown in Fig.~\ref{fig:zspan}. The white dot
labeled GW190728\rule{4pt}{.5pt}064510
corresponds  to the   best localized event among  the most likely lensed pair candidates quoted in \cite{LIGOScientific:2021izm}; in this case only the error 
$(\Delta d_L/d_L)_{\rm GW}$ has been included. The white dot labeled O4 corresponds to an example of an  event that could be  expected in the O4 run, assuming $(\Delta d_L/d_L)_{\rm GW}=20\%$ and 
$(\Delta d_L/d_L)_{\Delta t}=10\%$, and taking $z_s=0.4$, which is
 the average detection redshift for LIGO in O4 for  an equal mass BBH of $30\, {\rm M}_{\odot}$ (see Tab.~2 of  \cite{Chen:2017wpg}). The dotted lines  set the limit for the $1\sigma$ and $2\sigma$ exclusion of the GR value $\Xi_0=1$, for our fiducial value $\Xi_0=1.8$,  and when not adding random displacements for the measured distances as mentioned above. Given that the value $\Xi_0=1.8$ is already the highest currently obtained from a viable cosmological model, for a single quadruple lensed event the results  for $\Xi_0$ start to be interesting only  in the region of the plane below the $1\sigma$ line, and only below  the $2\sigma$ line one could start to have some hint for deviations from GR. As mentioned, for every second event, on average, the median measurement of $\Xi_0$ will be larger than $1.8$ due to noise. This has the effect of shifting the $1\sigma$ and $2\sigma$ lines upwards, which offers the chance of detecting modified gravitational wave propagation with higher significance than shown here.  Observe also that,
of course, with $N$ lensed events which give a  comparable  value of $\Delta\Xi_0/\Xi_0$,
the  combined error will be further reduced by a factor of order $1/\sqrt{N}$.

\begin{figure}[t]
\centering
\includegraphics[width=0.5\textwidth]{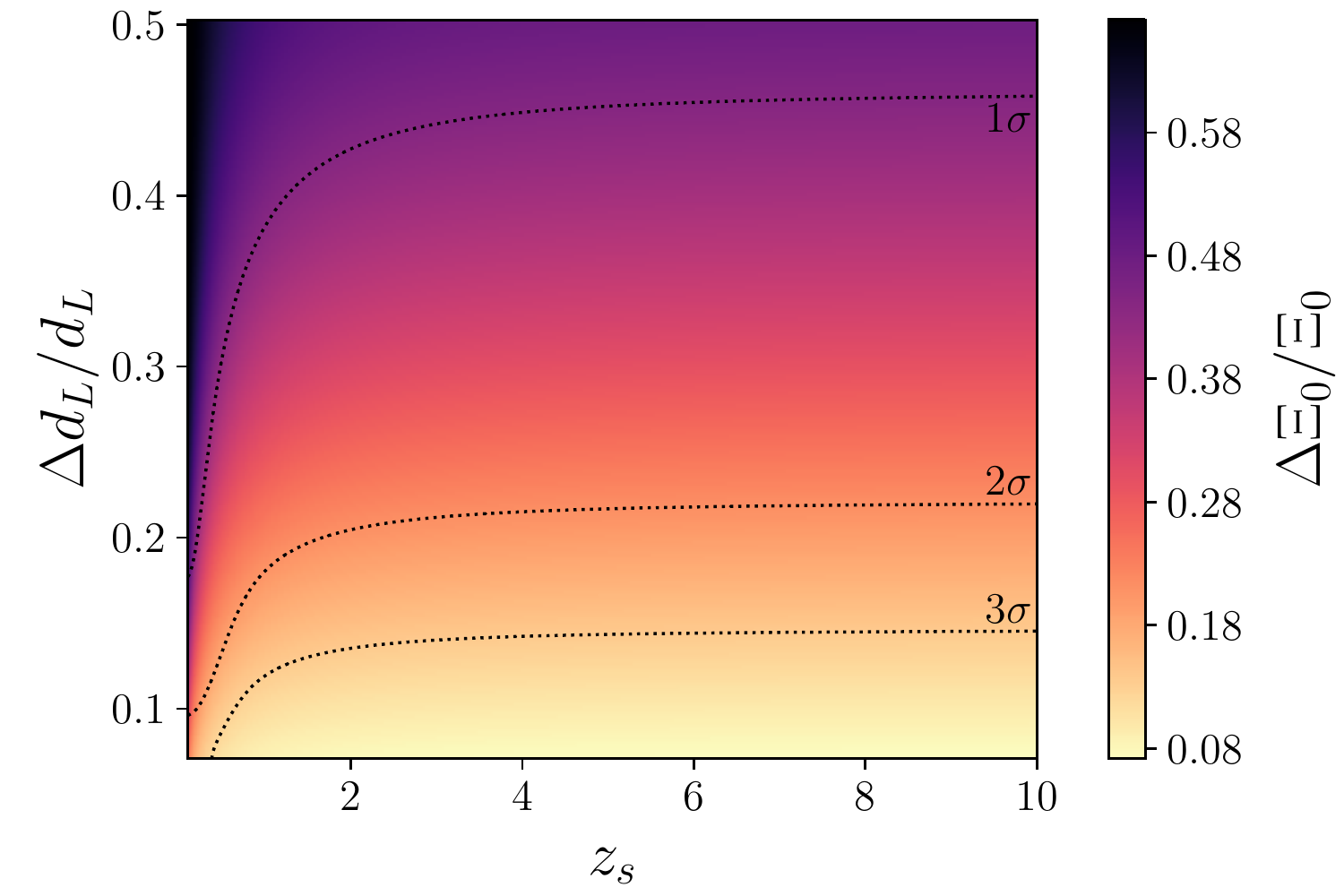}
\caption{As in Fig.~\ref{fig:mesh}, for a range of redshifts more appropriate to Einstein Telescope.} 
\label{fig:meshET}
\end{figure}

Fig.~\ref{fig:meshET} shows again $\Delta\Xi_0/\Xi_0$, for a   range of redshifts and of 
$\Delta d_L/d_L$ 
more appropriate to a 3G detectors such as the Einstein Telescope (which can actually even reach much higher redshifts, for which, however, a dedicated study of lensing reconstruction should be performed).
A first order of magnitude estimate for the value of $\Delta\Xi_0/\Xi_0$ that ET could reach from quadruply lensed events can be obtained observing that at ET, and even more with a  network involving 
another third-generation (3G) detector such as Cosmic Explorer,  the error  
$\Delta d_L /d_L$ will  likely be dominated by the lens reconstruction. If, as a first estimate,  we assume  $\Delta d_L /d_L\sim 25\%$, we see from Fig.~\ref{fig:mesh} that, at the large redshifts explored by 3G detectors, this corresponds to 
$\Delta\Xi_0/\Xi_0\sim 0.3$. Such a  30\% accuracy would already be sufficient to test the prediction $\Xi_0\simeq 1.8$ from nonlocal gravity with an interesting accuracy, with a single quadruply lensed event. Furthermore,  assuming, at ET,  100 strong lensing events per year \cite{Biesiada:2014kwa}, of which $6\%$ will be quadruply lensed~\cite{Li:2018prc}, in four years ET could collect ${\cal O}(25)$ quadruply lensed events, so the combined error would be of order $\Delta\Xi_0/\Xi_0\sim 0.3/\sqrt{25}=6\%$.

Note that  the case discussed at the beginning of this section, where $\dem$ is determined from the redshift $z_s$ assuming a value of $H_0$ and $\oma$, can also be included as a special case of Figs.~\ref{fig:mesh}
or \ref{fig:meshET}: simply, in this case the error $(\Delta d_L/d_L)_{{\Delta t}}$ in \eq{Deltatsudtot} is replaced by the error on $\dem(z_s)$ obtained from the redshift and the cosmological parameters. If the redshift is accurately known  from a spectroscopic follow-up, the error is dominated by the error on the assumed value of $H_0$ and $\oma$, which, using the {\em Planck\,} 2018 values, would be at the sub-percent level, so, in practice, $(\Delta d_L/d_L)_{{\Delta t}}$ would be negligible and 
 \eq{Deltatsudtot} will be replaced by 
$\Delta d_L/d_L\simeq  (\Delta d_L/d_L)_{\rm GW}$. Therefore, we will have a smaller value of $\Delta d_L/d_L$, at the price of having assumed a value for $H_0$.

\begin{figure}[t]
\centering
\includegraphics[width=0.4\textwidth]{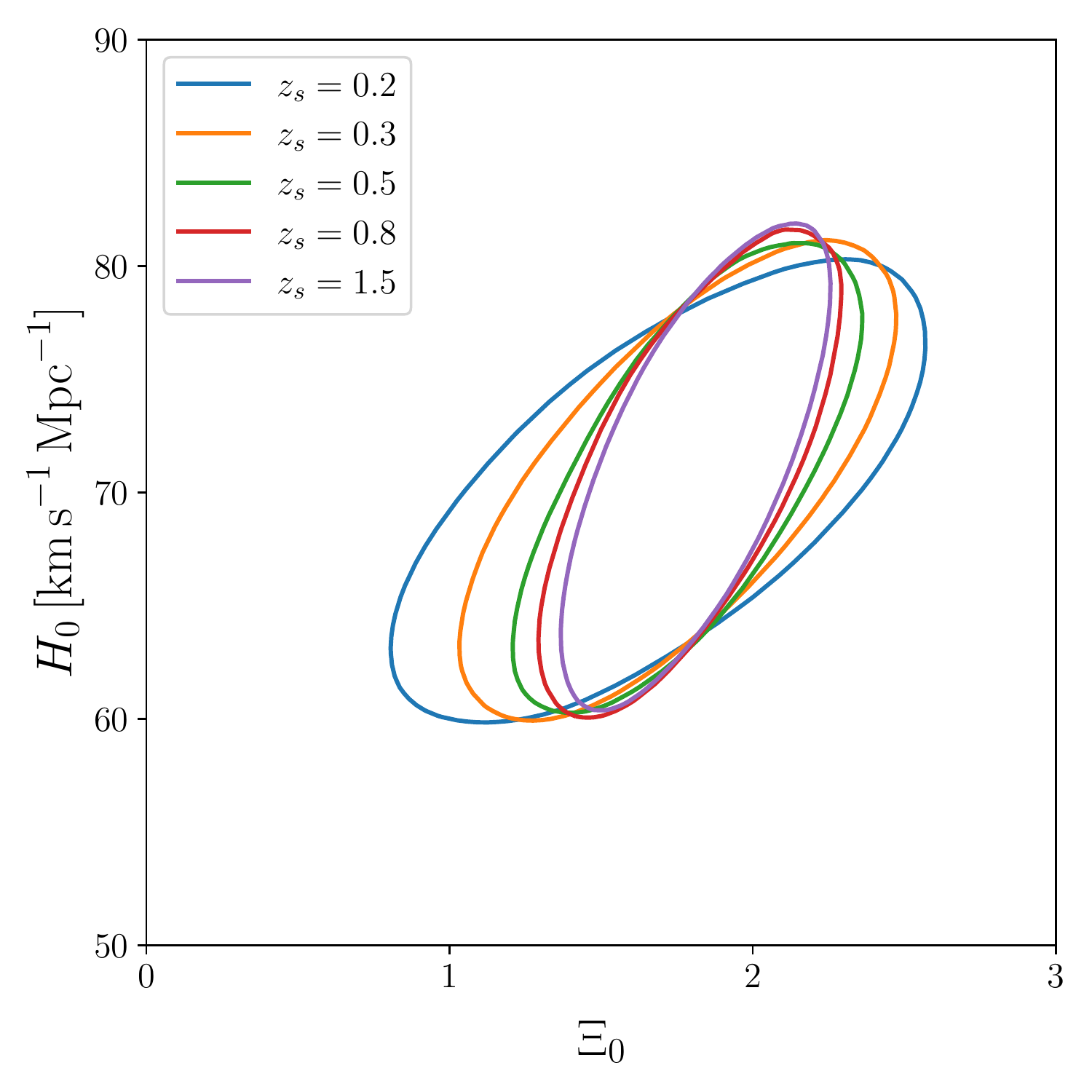}
\includegraphics[width=0.4\textwidth]{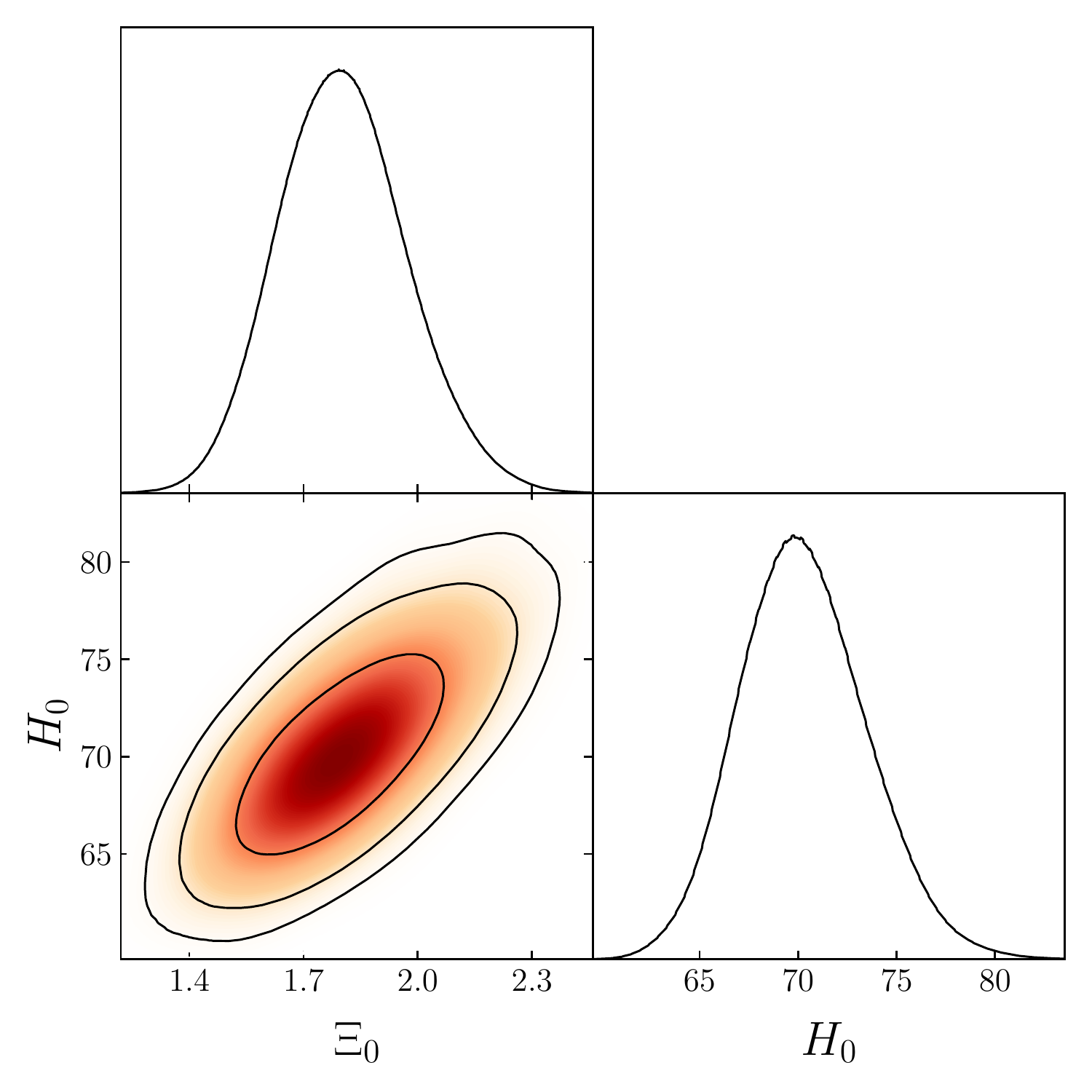}
\caption{Upper plot: $68\%$ contours of events at different redshifts,  assuming an error on $d_L$  of $10\%$. Lower plot:  the combination of all events of the upper plot. 
We use as 
fiducial values $H_0=70\, {\rm km}\, {\rm s}^{-1}\, {\rm Mpc}^{-1}$ and $\Xi_0=1.8$.
} 
\label{fig:2d}
\end{figure}

On the opposite side, one might want to explore what limits can be set on  both $H_0$ and $\Xi_0$ from a quadruply lensed event, without any prior on $H_0$. When opening up this two-dimensional parameter space, the degeneracy between $H_0$ and $\Xi_0$ from a $\dgw$ measurement with known redshift $z_s$ can be broken without an external prior on $H_0$, since the $\dem$ measurement constrains $H_0$. This is shown in the upper plot of
of~\ref{fig:2d}, where we show five events at different redshifts,  simulated with the same procedure  performed for obtaining   Figs.~\ref{fig:dLH0}-\ref{fig:mesh}. The plot also
shows that the $1\sigma$ contours are ellipses whose tilt changes with $z_s$, so the combination of events at different redshifts helps to break the degeneracy further. The lower plot shows  an example of the combined posteriors from the five events in the upper plot. From this example of simulated events, we get a combined result $\Xi_0=1.80^{+0.16}_{-0.18}$ and $H_0 = 70.3\pm 3.1$, so our fiducial values are recovered to an accuracy of about $9\%$ for $\Xi_0$ and $4\%$ for $H_0$. Note that, to avoid cluttering the plot, we are again neglecting to simulate the noise that would scatter these ellipses according to their size. In practice, the result we present here for 5 events is achieved, and even improved upon, by half of a set of 10 real events alone, while the remaining half of events will only contribute insignificantly to the constraint.


\section{Conclusions}\label{sect:conclusions}

The observation of strong lensing of GW events might already be possible for  2G detectors, and is expected to be common for 3G detectors. A fraction of these events (from $30\%$ at 2G to $6\%$ at 3G detectors) could be quadruply lensed. As discussed in \cite{Hannuksela:2020xor}, for these events it can be possible to localize the host galaxy even in the absence of an electromagnetic flare, and therefore these systems could be used as standard sirens.
In this paper we have explored the consequences in the context of modified gravity, in particular for modified gravitational wave propagation. We have pointed out that, in such a context, a quadruply lensed system allows the simultaneous determination of the electromagnetic luminosity distance and of the gravitational-wave luminosity distance of the source. Quite interestingly, with this method one can obtain limits on $\Xi_0$,  which is the most important parameter that enters in the parametrization (\ref{eq:fit}) of modified GW propagation, without imposing a prior on $H_0$, and one can also perform a joint  inference for $(H_0,\Xi_0)$.

Our main results are summarized in Fig.~\ref{fig:mesh}, that shows the relative error that can be obtained on $\Xi_0$,  as a function of the redshift of the source and of the error on the measurements of the electromagnetic and gravitational luminosity distances (combined in quadrature). Comparing with the largest prediction available from a viable cosmological model, which is 
$\Xi_0\simeq 1.8$ for  nonlocal gravity~\cite{Belgacem:2019lwx,Belgacem:2020pdz}, corresponding to a $80\%$ deviation from the GR value $\Xi_0=1$, we see that significant results might already be obtained by  2G detectors, while a third-generation detector such as ET, combining the events observed over a few years, might reach a very interesting accuracy, possibly  of order $6\%$, on $\Xi_0$.

\vspace{5mm}\noindent
{\bf Acknowledgments.} 
The work  of the authors is supported by the  Swiss National Science Foundation and  by the SwissMap National Center for Competence in Research. 


\bibliographystyle{utphys}
\bibliography{myrefs}

\end{document}